\documentclass[letters, usegraphix, useAMS, fleqn, usenatbib]{mnras}

\usepackage{amsmath}
\usepackage{amssymb}
\usepackage{booktabs}
\usepackage{xcolor}
\usepackage{graphicx}
\usepackage{dcolumn}
\usepackage{bm}
\usepackage{hyperref}
\hypersetup{colorlinks, allcolors={blue}}
\usepackage{soul}
\usepackage{siunitx}
\usepackage{multicol}

\usepackage{csquotes}
\usepackage{physics}
\usepackage{verbatim}
\usepackage{subcaption}
\usepackage{amsmath}
\usepackage{orcidlink}

\newcommand{\sjm}[1]{\textcolor{blue}{[SJM: #1]}}

\title[]{Realistic Equations of State Informing Neutron Star Post-Merger Gravitational-Wave Frequencies}

\author[Magnall et al.]{
\parbox{\textwidth}{
Spencer J. Magnall$^{1,2}$~\orcidlink{0009-0000-7037-1809},
Nilaksha Barman$^{3}$~\orcidlink{0009-0008-1220-359X},
Debarati Chatterjee$^{3}$~\orcidlink{0000-0002-0995-2329},
Paul D. Lasky$^{1,2}$~\orcidlink{0000-0003-3763-1386},
Simon Goode$^{1,2}$~\orcidlink{0000-0002-9575-5152}
}\vspace{0.4cm}\\
\parbox{\textwidth}{
$^{1}$School of Physics and Astronomy, Monash University, VIC 3800, Australia\\
$^{2}$OzGrav: The ARC Centre of Excellence for Gravitational-wave Discovery, Australia\\
$^{3}$Inter-University Centre for Astronomy and Astrophysics, Pune University Campus, Pune 411007, India
}}

\begin{document}
\maketitle

\begin{abstract}
Binary neutron star mergers are thought to produce hot, rapidly rotating neutron stars with masses that can far exceed their Tolman-Oppenheimer-Volkoff mass. The gravitational-wave emission from such remnants provides a unique opportunity to measure the nuclear equation of state at densities and temperatures not available to terrestrial experiments. Current detector design is informed by gravitational-wave signals from general relativistic hydrodynamics simulations of neutron star mergers, typically with hybrid thermal treatments for the equation of state, where a cold equation of state is modified by adding a thermal component.  
We use realistic equations of state based on the relativistic mean field model with consistent treatment of thermal effects to compute the distribution of  expected peak gravitational-wave frequencies. Marginalising over equation of state and progenitor neutron star masses, we show the peak frequency of emission ranges from $\sim2.5$ to 4 kHz. The width of this distribution suggests the need for broadband observatories with kHz sensitivity, and calls into question some of the so-called post-merger optimised configurations. We show the proposed KAGRA high-frequency design is well-suited to measuring post-merger remnants when compared to the KAGRA broadband design.
\end{abstract}

\section{\label{sec:intro}Introduction}

The observation of gravitational waves (GWs) from the binary neutron star merger GW170817~\citep{Abbott17_GW170817} provided a wealth of knowledge about the internal structure of neutron stars and cold dense nuclear matter at densities unable to be probed by terrestrial experiments~\citep{Lattimer_2021ann_rev}. Observations of neutron star masses and radii from radio and X-ray observations~\citep{NICER_2014,Riley_2019,Miller_2019,Riley_2021,Miller_2021} further inform our understanding of cold nuclear matter at many times nuclear saturation density. Neutron star mergers also have the potential to constrain the so-called ``hot'' equation of state (EoS) via observations of GWs from the hot 
($ \sim 100\, \rm MeV$), highly magnetised ($B \gtrsim 10^{15} \rm \, G$), and rapidly rotating nascent neutron star sometimes formed in the aftermath of a merger.     

Depending on mass, magnetic field strength, and thermal profile, remnants may either promptly collapse into a black hole, form a hypermassive neutron star supported against collapse by differential rotation and thermal gradients, form a supramassive neutron star supported against collapse by rigid rotation once the differential rotation is saturated, or form a stable neutron star (see \citealt{Sarin_Lasky} for a review). 
In either of the latter three scenarios, the merger remnant is expected to emit gravitational waves due to the excitation of a variety of oscillation modes defined by their restoring force (e.g pressure, gravity, rotational, etc.). The largest-amplitude emission is expected to be from the fundamental pressure ($f$-) mode excited during the merger. As $f$-mode frequencies are sensitive to the internal composition, the post-merger signal of a binary neutron star merger provides a unique opportunity to probe the hot nuclear equation of state. 

The $f$-mode frequency of a non-rotating neutron star scales with the square root of average mass density~\citep{Andersson_Kokkotas_1998mnras} . This relationship has now been modified to include rigid rotation, a large range of potential equations of state, and thermal effects for hot equations of state~\citep{Doneva_2013,2021Pradhan,Pradhan_Chatterjee_2022prc,Barman_2025,Barman_Chatterjee_2025arxiv}. These quasi-universal relations can help constrain dense matter properties from GW data using asteroseismology, with the most likely scenario being the observation of oscillation modes from post-merger remnants. Such signals can provide EoS constraints via peak frequency measurements~\citep[e.g,][]{2023_bayes_stack,2025Mitra, 2025Miravet_tenes} and enable remanent classification via morphology-independent methods~\citep[e.g,][]{2023Tringali}.

Searches following the detection of GWs from the inspiral portion of neutron star merger events GW170817 and GW190425 did not find any detectable post-merger signals~\citep{2017GW170817_PM_search,GW190425}.  Numerical simulations and theoretical predictions show that they are not expected to be measurable with current detector sensitivities~\citep{Maggiore_2020,Abac_2025}. First detection of post-merger GWs may rely on future upgrades of current observatories to e.g., the so-called A$\sharp$ design upgrade to LIGO~\citep{Asharp}, or even third-generation observatories Einstein Telescope~\citep{EinsteinTelescope,2011_Hild_ET} or Cosmic Explorer~\citep{CosmicExplorer,2019_CosmicExplorer}.  On the other hand, high-frequency GW detectors such as the proposed Neutron Star Extreme Matter Observatory (NEMO)~\citep{ackley20}, or the proposed high-frequency upgrade to the KAmioka GRAvitational wave detector~\citep[KAGRA;][]{KAGRA,2021Akutsu,2025KAGRA_HF}, aim to target post-merger remnants by enhancing narrowband sensitivity in the range of $\sim 1$---4 kHz.  

Narrowband configurations with a bandwidth of a few hundred Hertz (utilising detuned signal recycling cavities) can target post-merger signals more effectively, especially since the chance of observing a post-merger remnant with a high signal-to-noise ratio is low. In this case, identifying theoretically the peak emission frequency becomes essential for optimising detector performance. 
However, predicting the expected gravitational-wave emission of nascent neutron stars is challenging as the remnant is expected to be out of beta equilibrium and 
undergoing rapid and significant thermal evolution. Consequently, we must use realistic finite temperature EoSs to improve the accuracy of post-merger frequency predictions. 

Current expectations for post-merger frequencies consider predictions based on merger simulations using limited numbers of parametric EoSs~\citep[e.g.,][]{Oechslin_Janka_2007prl,Sekiguchi_Kiuchi_2011prl,Bauswein_Baugmarte_2013prl,Takami_2015,Radice_Bernuzzi_2017apjl}. These simulations typically employ a ``hybrid" thermal treatment, i.e., a cold neutron star EoS supplemented by an ideal gas thermal contribution using a gamma ($\Gamma$) function, with a fixed value of $\Gamma\in(1.5,2)$~\citep{Weih_2020}. However, recent work shows that this formalism fails when non-nucleonic degrees of freedom such as hyperons are included, which are expected to be present at merger conditions of high temperatures and densities~\citep[e.g.,][]{Raduta_Nacu_2021epja,Raduta_2022epja,Kochankovski_Ramos_2022mnras,Blacker_Kochankovski_2024prd,Barman_Chatterjee_2025arxiv}. Ideally, simulations should implement a realistic finite temperature EoS instead of the ``gamma-law" prescription. However, this is challenging as the EoS tables are three dimensional (temperature $T$, baryon density $n_b$ and charge fraction $Y_Q$) and cover a large parameter space: $0 < T < 100 \, \rm MeV$, $ 10^{-10} < n_B/n_{sat} < 6.0 $ (where, $n_{sat}$ is the nuclear saturation density), $0.01 < Y_Q < 0.6$.

Linking this all together, it is clear that understanding potential GW peak frequencies, and hence optimising observatory design, relies on calculations of $f$-mode frequencies with large sets of realistic EoSs that include thermal effects. In this \textit{Letter}, we construct a large set of possible EoSs using a phenomenological relativistic mean field model informed by theoretical, experimental and astrophysical constraints. We quantify the impact of finite temperature effects on the peak GW frequency for the post-merger phase of a binary neutron star merger. We show that the inclusion of finite temperature effects most favours a detector with peak sensitivity at $\sim 3000 \, \rm Hz$. We compare detuned post-merger optimised interferometers to their broadband configurations and quantify the improvement in signal-to-noise ratio. 
We show that scientific output is maximised by using a high-frequency optimised detector detuned at $\sim 3 \,\rm kHz$. However, we also caution that the width of the peak frequency distribution should be taken into account when considering the narrowband nature of potential detuned observatories.

\section{Methods}\label{sec:methods}

\subsection{Equation of state construction}
The EoS of dense matter depends on the underlying saturation properties of nuclear matter. 
These include nuclear saturation density ($n_{sat}$), energy per baryon at saturation of symmetric nuclear matter ($E_{sat}$), incompressibility at saturation of symmetric nuclear matter ($K_{sat}$), symmetry energy at saturation ($J_{sym}$), slope of symmetry energy at saturation ($L_{sym}$) and effective nucleon mass ($m^*/m$) and higher order parameters.
However, large uncertainties on nuclear parameters  produces large uncertainties in the nuclear EoS~\citep{Lattimer_Prakash_2004science,Lattimer_2012annurev-nucl,Glendenning,Schaffner-Bielich_2020,Haensel_Potekhin_2007,Rezzolla_Pizzochero_2018,Baym_Hatsuda_2018rpp,Burgio_Schulze_2021ppnp}. In Table~\ref{tab:prior}, we list the uncertainty ranges of the empirical nuclear parameters. 

\cite{Chen_Piekarewicz_2014prc} introduced a non-linear relativistic mean field model (NL-RMF) that describes dense nuclear matter through baryon interactions mediated by scalar $\sigma$, vector $\omega$ and isovector $\rho$ mesons.  By fitting the coupling strength of mesons to baryons, the model reproduces the desired nuclear saturation properties~\citep{Hornick_Tolos_2018prc}. 
Subsequent works~\citep{Ghosh_Chatterjee_2022epja,Ghosh_Pradhan_2022_fspas,Maiti_Chatterjee_2024arxiv} used the NL-RMF model within a Bayesian scheme to constrain the parameter space of cold neutron star EoSs using information from multi-disciplinary physics, i.e., microscopic calculations from Chiral Effective Field Theory ($\chi$EFT) at low density (0.5 - 1.4~$n_{sat}$)~\citep{Drischler_Carbone_2016prc} and astrophysical observations at high density (beyond 3~$n_{sat}$).
Later work~\citep{Barman_2025,Barman_Chatterjee_2025arxiv} extended this formalism to include finite temperature effects. We adopt here the same formalism, as it allows us to consistently construct both zero-temperature and finite-temperature EoSs that satisfy state-of-the-art multi-disciplinary physics constraints. 

\begin{figure}
    \centering
    \includegraphics[width=\linewidth]{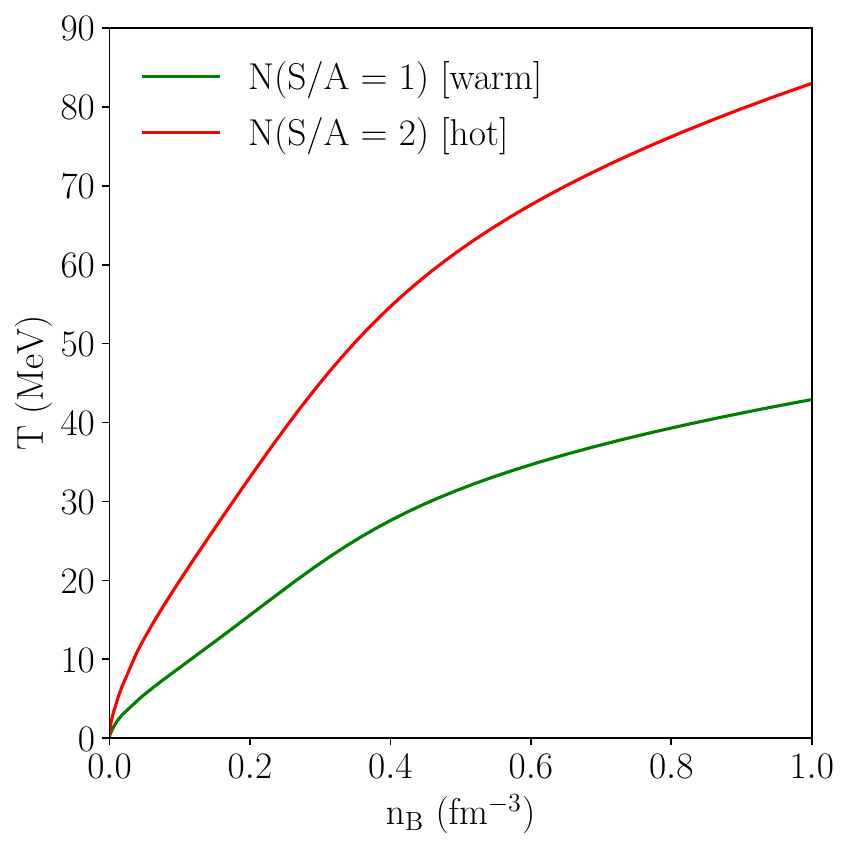}
    \caption{Temperature as a function of baryon density for nucleonic matter for warm ($S/A=1$) and hot ($S/A=2$) neutron star  configurations. 
    }
    \label{fig:T_vs_nb_S_1_2}
    
\end{figure}

Numerical simulations of binary neutron star post-merger remnants, show that temperatures inside the remnants reach $\sim20-80$ MeV (e.g., see Fig. 1 of~\citealt{Sekiguchi_Kiuchi_2011prl}), and that the entropy per baryon of the degenerate matter is $S/A\sim1-2$~\citep{Most_Motornenko_2023prd}. In this work, we analyse three thermodynamic configurations: (a) cold; $T=0$, (b) warm; $S/A=1$ and (c) hot; $S/A=2$. 
In Fig.~\ref{fig:T_vs_nb_S_1_2}, we show the temperature profile as a function of baryon density for the warm and the hot configurations for one set of nuclear parametrisation (see the ``Fixed'' configuration in Table~\ref{tab:prior}).
The temperature profiles demonstrate that our EoSs reach temperatures relevant for conditions inside post-merger remnants. We note that this is an approximate thermal treatment of the hot remnant in which it has a temperature gradient given by fixed entropy per baryon, and we ignore gradients in the entropy per baryon.

\begin{table*}
    \centering
    \caption{Parameter sets of nuclear empirical properties used in this study. The top row is the ``Fixed" set of nuclear parameters within their allowed uncertainty range used for the study of thermal effects, while the bottom row considers the entire range for the Bayesian analysis.}
    \label{tab:prior}
    \begin{tabular}{lcccccc}
        \toprule
        \toprule
         &$n_{sat}$ & $E_{sat}$ & $K_{sat}$ & $J_{sym}$ & $L_{sym}$ & $m^*/m$  \\
         &($\rm{fm^{-3}}$) & (MeV) & (MeV) & (MeV) & (MeV) &   \\
         \midrule
        Fixed & 0.15 & -16 & 240 & 32 & 60 & 0.65 \\
        \midrule
        Range & 0.14 to 0.17 & -16$\pm$0.2 & 200 to 300 & 28 to 34 & 40 to 70 & 0.55 to 0.75 \\
        \bottomrule
        \bottomrule
    \end{tabular}
\end{table*}

 For nucleonic matter, we uniformly draw from nuclear parameters within the uncertainty range allowed by current experiments as given in Table~\ref{tab:prior}. 
 We further constrain the nuclear parameters using  information from $\chi$EFT~\citep{Drischler_Carbone_2016prc} and astrophysical observations~\citep{GW170817,Eemeli_Tyler_2018prl,Most_Weih_2018prl,Tong_Zhao_2020prc}; and combined set of parameters that yields EoSs inconsistent with these constraints are discarded. 
After applying these constraints, we generate $\sim3300$ nuclear parameter sets and use them to construct EoSs for cold ($T=0$), warm ($S/A=1$), and hot ($S/A=2$) post-merger remnant configurations.
The mass-radius relations for non-rotating neutron stars using our EoS sets are shown in Appendix~\ref{sec:mass_radius_app}.

\subsection{$f_{peak}$ calculations}

We estimate of the dominant GW emission frequency $f_{\rm peak}$ by calculating the emission of the co-rotating $l=|m|=2$ $f$-mode. 
This ideally requires numerical simulations considering a wide range of hot EoSs, however this is infeasible due to the high computational costs. Therefore, in order to get an estimate of the impact of thermal effects on peak GW frequencies, we make perform semi-analytic calculations utilising several approximations.

Our calculations closely follow the arguments of \cite{2020Chakravarti}. Firstly, we neglect thermal evolution of the remnant and  assume that the temperature profile of the EoS is roughly the profile of the post merger remnant (or at least has the dominant influence on structure).
Secondly, ignoring differential rotation, we assume that the hypermassive neutron star is mostly uniformly rotating, noting this is a reasonable approximation for the core, but not necessarily the outer layers~\citep[e.g.,][]{2017Uryu}. We then utilise the relations of \citealt{Doneva_2013} to calculate the co-rotating $f$-modes frequencies $\sigma_{\rm corot}$ from the static masses and radii
\begin{equation}
\label{eq:doneva_corot}
    \frac{\sigma_{\rm corot}}{\sigma_0} = 1 - 0.235 \left(\frac{\Omega}{\Omega_k}\right) - 0.358 \left(\frac{\Omega}{\Omega_k}\right)^2, 
\end{equation}
where $\sigma_0$ is the oscillation frequency of the $f$-mode for the non-rotating case, $\Omega$ is the hypermassive neutron star rotation frequency and $\Omega_k$ is the break up frequency of the neutron star (Kepler frequency). We approximate this Kepler frequency using the relation derived in \cite{Doneva_2013}
\begin{equation}
    \frac{1}{2\pi}\Omega_k\,[{\rm kHz}] = 1.716 \ \left({\frac{M_0}{1.4 \ M_\odot}}\right)^{1/2}  \left(\frac{R_0}{10 \ \rm km}\right)^{-3/2} -0.189,
\end{equation}
where $M_0$ and $R_0$ are the mass and radius of the non-rotating model, respectively. 
The mass scaling with rotation frequency of the neutron star  is given by,
\begin{equation}\label{eq:M_rot}
    \frac{M}{M_0} = 0.991 + 9.36\times10^{-3} \exp\left(3.28\frac{\Omega}{\Omega_K}\right)~,
\end{equation}
where $M$ is the mass of the rotating neutron star . The frequency for the non-rotating $l= |m| = 2$ is given by 
\begin{equation}\label{eq:sigma_0}
    \frac{1}{2\pi } \sigma_0\,[{\rm kHz}] = 1.562 + 1.151 \left(\frac{M_0}{1.4\, M_\odot}\right)^{1/2} \left(\frac{R_0}{10 \,\rm km}\right)^{-3/2}.
\end{equation}
Assuming a remnant rotating at the Kepler frequency and substituting $\Omega_k$ and $\sigma_0$ into Equation~\ref{eq:doneva_corot}, we then convert from the co-rotating frame to the inertial frame 
\begin{equation}\label{eq:sigma_inert}
    \sigma_{\rm inert} = \sigma_{corot} - m\Omega.
\end{equation}

We do not expect our calculations of frequency to be accurate when compared to numerical-relativity simulations. Rather, we need them to be \textit{accurate enough} to inform future detector design. The use of the Cowling approximation in the relations of \cite{Doneva_2013} implies errors of approximately $10-30 \%$ when compared to a fully general-relativistic treatment~\citep{Pradhan_Chatterjee_2022prc}. Similarly, our approximate treatment of the thermal profile and neglecting differential rotation will also introduce errors in the calculated peak frequencies. 

For each EoS, we compute the static NS mass radius configuration by solving the TOV equations. We estimate the non-rotating $l=2$ $f$-mode frequency by using Equation~\ref{eq:sigma_0}. We then assume a uniform distribution of non-rotating NS mass configurations upto its maximum non-rotating mass. Adding Kepler rotation, the mass of these NS configurations increases by a factor of $\sim24\%$ as can be deduced from Equation~\ref{eq:M_rot}. Thus for each EoSs, we get a uniform distribution of rapidly rotating NS configurations that spans the gravitational mass of the post-merger remnants in the range $\sim2.2 M_\odot-1.24 M_{TOV}$. Finally, we estimate the peak GW frequency for a given post-merger mass configuration as $f$-mode frequency observed by an inertial observer. It is given by Equation~\ref{eq:sigma_inert}.


\begin{figure}
    \centering
    \includegraphics[width=\linewidth]{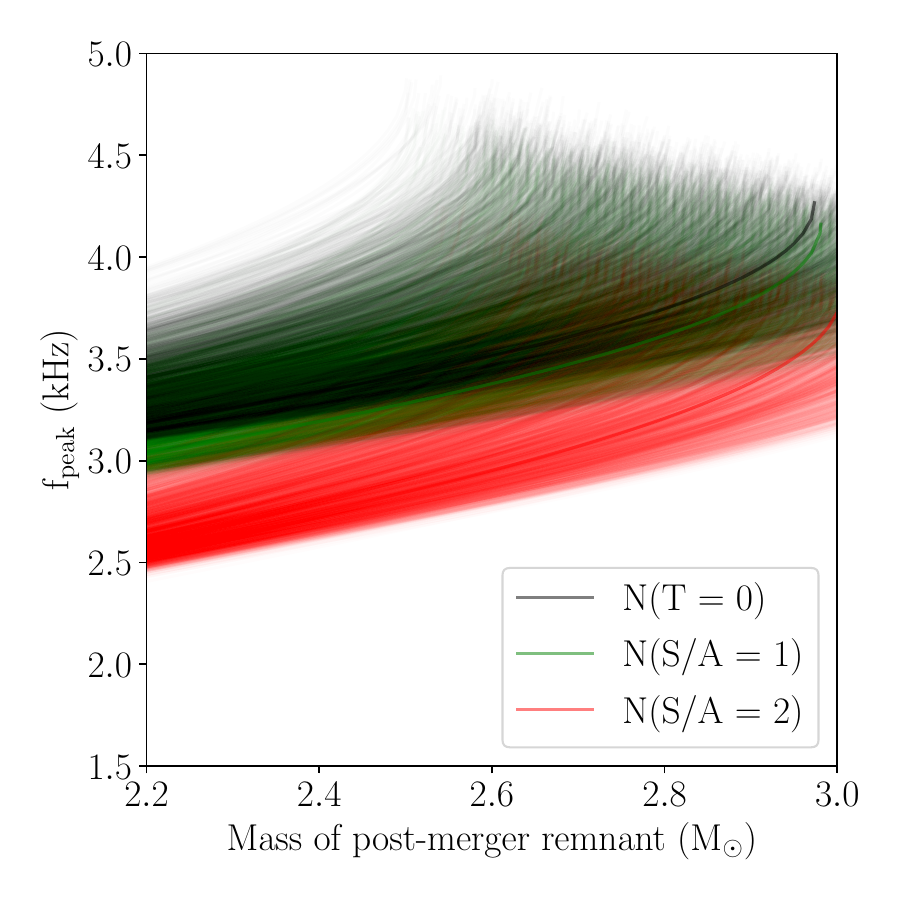}
    \caption{Peak GW frequency as a function of mass of post-merger remnant at Kepler rotation for cold ($T=0$; grey), warm ($S/A=1$; green) and hot ($S/A=2$; red) EoSs.}
    \label{fig:f_peak_cold_hot}
\end{figure}

\begin{figure}
    \centering
    \includegraphics[width=\linewidth]{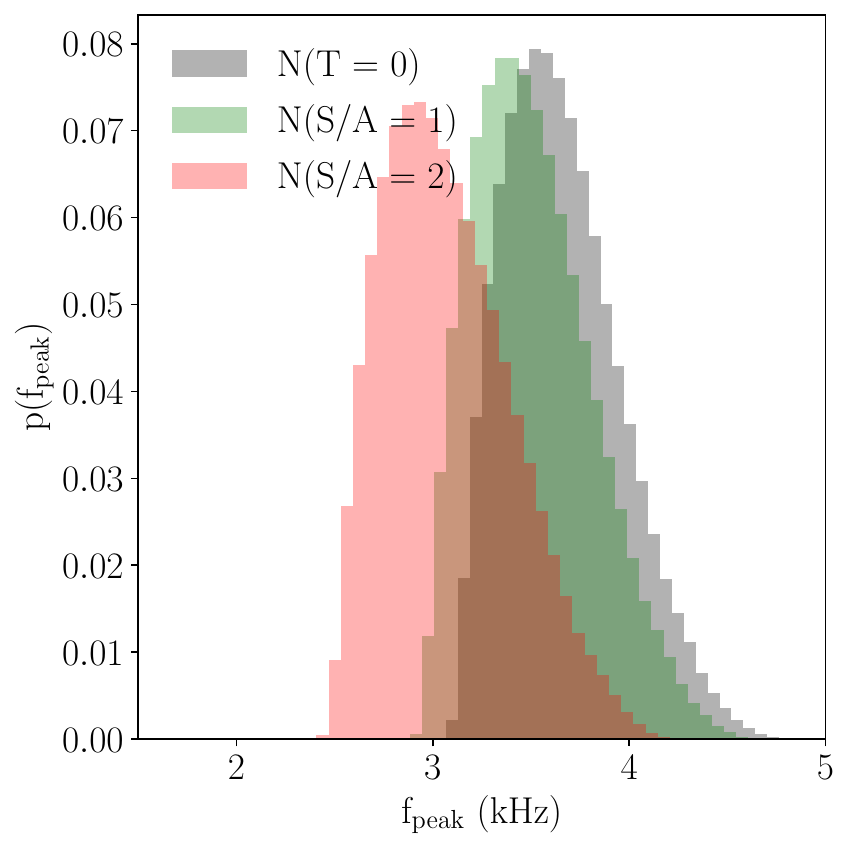}
    \caption{Peak GW frequency distributions with zero and finite temperature EoSs for all post-merger remnant masses. Shown are results for cold ($T=0$; grey), warm ($S/A=1$; green) and hot ($S/A=2$; red) remnants.}
\label{fig:f_peak_hot_cold_posterior_total}
\end{figure}

\subsection{Post-Merger Optimised Detector Configurations}

To quantify the benefit (or detriment) of post-merger optimised interferometers,
we calculate the signal-to-noise ratio of a post-merger remnant for both broadband and optimised configurations. 
We consider the proposed high-frequency upgrade to KAGRA~\citep{KAGRA,2025KAGRA_HF} as our baseline set of configurations. For the optimised detectors, we use the 2 kHz and 3 kHz configurations of KAGRA HF~\footnote{KAGRA HF noise sensitivity curves can be found at \href{https://gwdoc.icrr.u-tokyo.ac.jp/cgi-bin/DocDB/ShowDocument?docid=16182}{JGW-T2416182-v3}}. 

As we only consider the dominant $f$-mode peak for each post-merger remnant, our GW signal is described by a time-domain model made of an exponentially-damped sinusoid where the plus polarisation of gravitational-wave strain is given by  
\begin{equation}\label{eq:damped_sine}
    h_{+}(t) = A\exp\left({-\frac{t}{t_{\rm damp}}}\right)\cos(2\pi f_0 t), 
\end{equation}
where $A$ is the amplitude, $t_{\rm damp}$ is the damping time, and $f_0$ is the peak frequency. 
We take the amplitude to be $A = 10^{-22}$ and set the damping time to be $t_{\rm damp} =0.025 \, \rm s$. We note that the amplitude and damping times of our signal should have an EoS dependence. However, calculating the correct damping times and amplitudes is computationally expensive, and we do not expect it to be a dominant source of error compared to other approximations made during the generation of our EoSs and peak frequencies.
We construct the cross polarisation strain by introducing a phase offset of $\pi/2$ to the plus polarisation time series in Equation~\ref{eq:damped_sine}. 

The optimal signal-to-noise ratio for a signal in a single detector is calculated via 
\begin{equation}
    \rho^2 = 4 \int^{\infty}_{0} \frac{|h(f)|^2}{S_n(f)}df,
\end{equation}
where $S_n(f)$ is the detector's noise power spectral density. 

We use \textsc{bilby}~\citep{Bilby_Ashton,Bilby_Romero-Shaw} to inject our signal and calculate the optimal signal-to-noise ratio. 
We construct a signal injection set of $\mathcal{O}(10^5)$ binaries for each cold, warm, and hot EoS set, which is averaged over all sky positions, and contains post-merger remnant masses drawn uniformly between $2.2M_\odot$ and $1.24 M_{TOV}$. We  inject these signal sets into both the broadband and optimised detectors, and construct a normalised signal-to-noise ratio via 
\begin{equation}
    \rho_{\rm norm} = \frac{\rho_{\rm PMO}}{\rho_{\rm BB}},\label{eq:normalisedSNR}
\end{equation}
where $\rho_{\rm PMO}$ is the signal-to-noise ratio of the injection in the post-merger optimised detector, and $\rho_{\rm BB}$ is the signal-to-noise ratio of the injection in the broadband configuration.

\section{Results}
The peak GW frequency as a function of remnant mass for all cold (grey curves), warm (green curves), and hot (red curves) EoSs is shown in Fig.~\ref{fig:f_peak_cold_hot}. We see that the peak GW frequency increases with increase of the mass of the merger remnant.

In Fig.~\ref{fig:f_peak_hot_cold_posterior_total}, we show the peak GW frequency distribution across a wide range of post-merger masses for the cold and hot EoSs considered. 
We find that the peak frequency decreases as the entropy of the star increases. 
This result is not unexpected as intuitively, finite temperature effects introduce a thermal pressure that ``puffs out" a neutron star decreasing its compactness. As the compactness is directly related to the $f$-mode frequency, a decrease in compactness leads to a systematic shift of the median of the peak frequency distribution.
We find that the median and 90\% credible intervals for the distributions of $f_{peak}$ for the cold ($T=0$), warm ($S/A=1$), and hot ($S/A=2$) equations of state are ${3627}^{+581}_{-389} \rm~Hz$, ${3476}^{+591}_{-397}\rm~Hz$, and ${3039}^{+633}_{-424}\rm~Hz$ respectively.

\begin{figure*}
    \centering
    \includegraphics[width=0.9\linewidth]{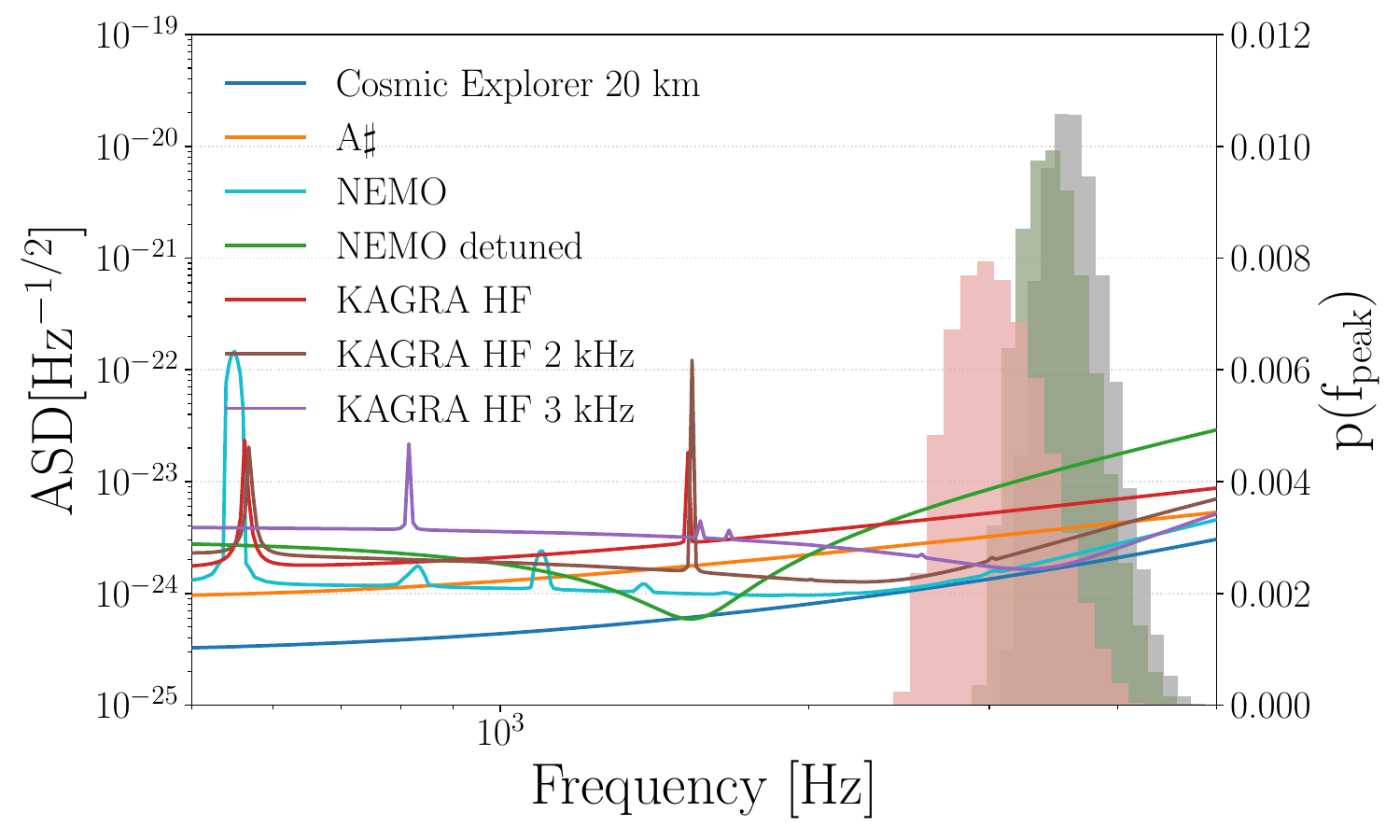}
    \caption{Distributions for the peak frequency of a post-merger remnant of a binary neutron star merger for cold (grey), warm (green) and hot (red) EoSs compared to the sensitivity curves of a 20 km Cosmic Explorer (blue), LIGO at A$\sharp$ sensitivity (orange), and various configurations of NEMO (blue, green) and a high frequency KAGRA detector (red, brown, purple).} 
    \label{fig:fpeak_CI_pos}
\end{figure*}

In Fig.~\ref{fig:fpeak_CI_pos}, we show the amplitude spectral density curves for the proposed interferometers NEMO (cyan curve), the $20 \ \rm km$ layout of CE (blue curve), LIGO at A$\sharp$ sensitivity (orange), and the high frequency upgrade to KAGRA (red curve). 
These curves are accompanied by proposed detuned or post-merger optimised configurations, a squeezed NEMO (green curve), and high frequency KAGRA configurations most sensitive at 2 kHz and 3 kHz (brown and purple curves, respectively). Overlayed are the peak-frequency distributions from Fig.~\ref{fig:f_peak_hot_cold_posterior_total}. We see that all EoS are best covered via the high frequency KAGRA detector at 3 kHz, with the standard NEMO configuration and the 20 km configuration of CE having similar sensitivity in the expected $f_{peak}$ frequency range. 

\begin{figure*}
    \centering
    \includegraphics[width=0.8\linewidth]{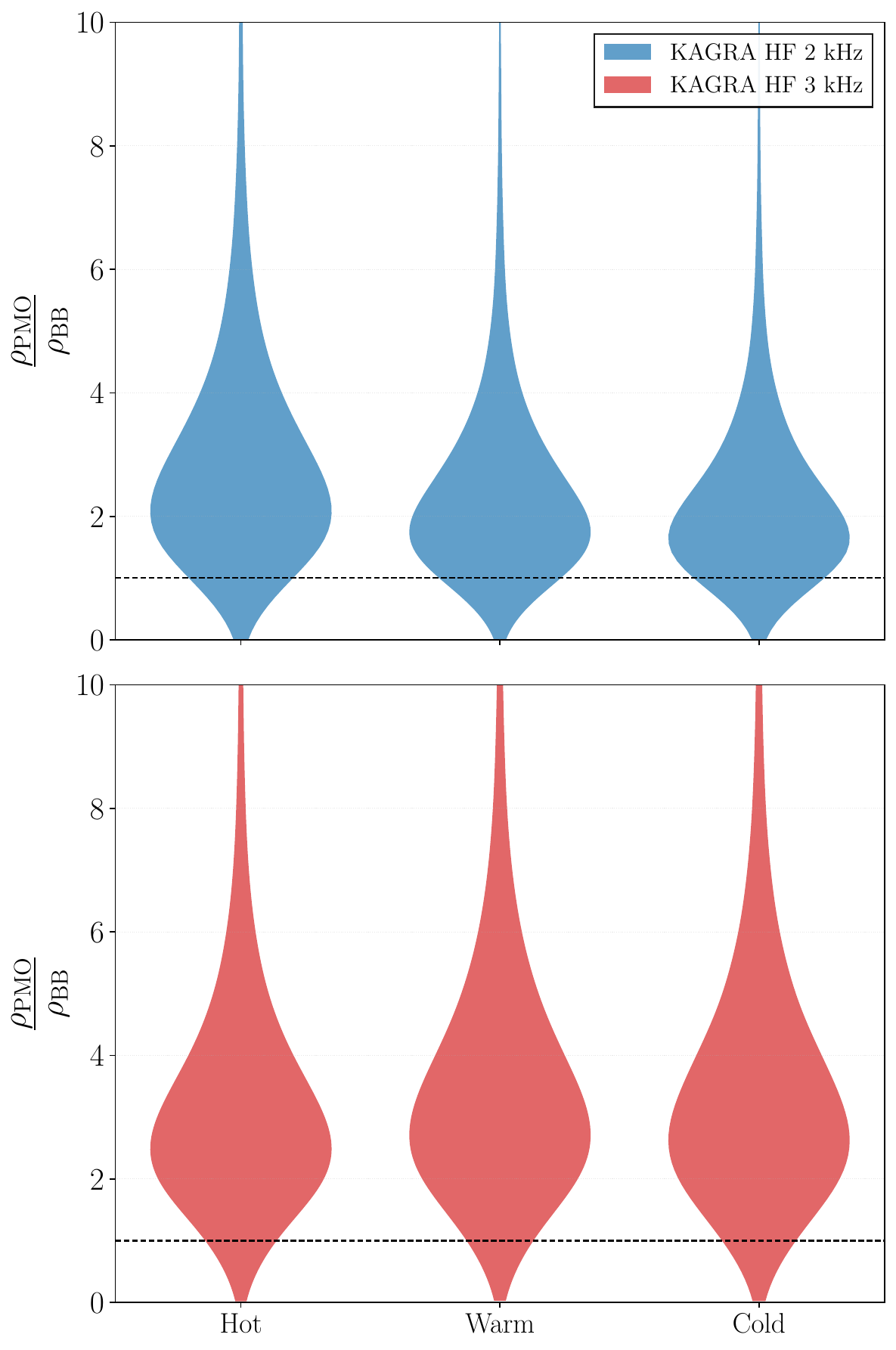}
    \caption{Violin plots of expected signal-to-noise ratios of post-merger remnants for our EoSs. The signal-to-noise ratios for the post-merger optimised detectors are normalised against the standard or `broadband' configuration of the detector. 
    The top panel corresponds to a 2 kHz configuration of the KAGRA HF detector whereas the bottom panel corresponds to a 3 kHz configuration of the KAGRA HF detector.
    Each panel contains three violin plots representing (from left to right) the hot ($S/A=2$),warm ($S/A=1$), and cold ($S/A=0$) equations of state. The black dashed line represents a signal-to-noise ratio of 1, i.e, exactly the same detector response from the broadband and optimised detectors.   }
    \label{fig:violin_plot}
\end{figure*}

Figure \ref{fig:violin_plot} shows violin plots of normalised signal-to-noise ratios (Eq.~\ref{eq:normalisedSNR}) for hot, warm and cold EoSs for the 2 kHz (top panel), and 3 kHz (bottom panel) configurations of KAGRA HF. The black dashed-line in each plot corresponds to a signal-to-noise ratio of unity. We see that across all EoS configurations, the 3 kHz configuration is most effective, offering on average an signal-to-noise ratio increase of approximately 2.5 times the standard configuration. However, the 2 kHz configuration still offers an improvement (on average) of 1.8-2 times the broadband configuration, with the improvement decreasing as EoS temperature decreases.

Throughout this work, we construct a post-merger signal model by focusing on the dominant $l=|m|=2$ fundamental mode emission only. Hydrodynamical simulations have shown that in addition to the fundamental emission there are also  at least three additional peaks in the spectrum of the post-merger gravitational wave emission 
(e.g., \citealt{2011Stergioulas,2015Bauswein,2016Rezzolla_Takami}).
These additional peaks are related to additional mode couplings and a spiral deformation of the post-merger remnant. 
Ideally, our signal model should consider these additional peaks, and in neglecting them we expect the optimal signal-to-noise ratio to be slightly lower.

\section{Conclusion}

In this work, we construct a large number of equations of state within a non-linear relativistic mean field model informed by $\chi$EFT and astrophysical observations. We consider cold ($T=0$), warm ($S/A=1$) and hot ($S/A=2$) thermodynamic conditions to describe a population of binary post-merger remnants across a range of gravitational masses.  We model the post-merger remnant as a rapidly rotating neutron star with Kepler rotation. These configurations emit gravitational waves that result in a post-merger frequency peak $f_{peak}$ corresponding to the fundamental quadrupolar oscillation mode.

We show that the inclusion of finite-temperature effects leads to a reduction of $f_{peak}$ by $\sim200-600$ Hz due to the increase of the size of the rotating star. We consider sensitivity curves of different configurations of the KAGRA HF detector and study the relative improvement in signal-to-noise ratios for post-merger optimised configurations compared to the relative broadband sensitivity. 
We find that the 3 kHz configuration of the KAGRA HF detector shows the highest improvement in the signal-to-noise ratio for post-merger physics. 

We make several approximations throughout the course of this work, the first of which is the assumption that our non-rotating hot neutron stars are well approximated by substituting an entropy per baryon of $S/A=1$ or $S/A=2$, which gives a thermal profile as seen in Figure~\ref{fig:T_vs_nb_S_1_2}. Snapshots from simulations of proto-neutron stars and of neutron star mergers show that this profile is accurate at least for a few cycles. However as the nascent neutron star is highly dynamical we expect the thermal profile to evolve on timescales relevant to the gravitational-wave emission timescale.

Secondly, rather than performing full calculations using general relativity, we use the relations of \cite{Doneva_2013} that utilise the Cowling approximation. We expect this to incur at least 30\% error, although expect this error to be consistent across equations of state. 
The \citeauthor{Doneva_2013} relations also assume uniform rotation and are calculated using zero-temperature equations of state. 

We construct equations of state under the assumption that they are purely nucleonic. However, many nuclear theories predict that neutron stars with masses we examine in this work should contain a large fraction of hyperonic or quark matter. The addition of new degrees of freedom softens the EoS and increases $f$-mode frequencies. A recent study~\citep{Barman_Chatterjee_2025arxiv} shows that hot hyperonic EoSs can sustain a maximum mass of only $\sim2.2M_{\odot}$. Hence uniform rotation alone is not sufficient to reach a gravitational remnant mass up to $\sim3M_{\odot}$ in such cases. Ideally we should include realistic differential rotation profiles that resemble conditions inside post-merger remnants and support higher masses across all kinds of EoSs. We leave a detailed study of the impact of differential rotation on post-merger emission to future works.

\section{Acknowledgements}
D.C. acknowledges support from the George Southgate fellowship and the kind hospitality at Monash University, where this work was conceptualised. D.C. thanks Chayan Chatterjee and Anarya Ray for helpful discussions. SJM, SG and PL thank Vaishali Adya and Bram  Slagmolen for helpful discussions. N.B. thanks Pranjal Tambe for providing useful insights regarding thermodynamic conditions in post-merger scenarios. 
SJM receives support from the Australian
Government Research Training Program.
SJM, PDL, and SRG are supported through Australian Research Council (ARC) Centres of Excellence CE170100004 and CE230900016, Discovery Projects DP220101610 and DP230103088, and LIEF Project LE210100002.
We also acknowledge computational support through the Ngarrgu Tindebeek / OzSTAR Australian national facility at Swinburne University
of Technology.

\bibliographystyle{mnras}
\bibliography{Paper}

\appendix


\section{Mass-Radius relations}\label{sec:mass_radius_app}
\begin{figure}
     \centering
     \includegraphics[width=\linewidth]{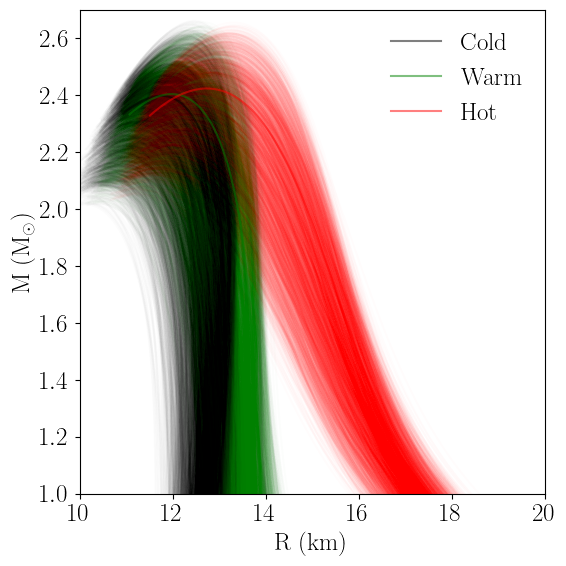}
     \caption{Gravitational mass as a function of radius for non-rotating neutron star  configurations for the cold $(T=0)$, warm $(S/A=1)$ and hot $(S/A=2)$ used in this work.}
    \label{fig:mr_hot_cold}
\end{figure}

 Fig.~\ref{fig:mr_hot_cold} shows the mass-radius posterior plot for non-rotating neutron star configurations corresponding to the cold, warm and hot EoSs used in this work in. We can see that thermodynamic profile with $S/A=1$ does not drastically alter the size of the star from its cold counterpart. Consequently, the resultant change in  $f$-mode frequencies is small. This is also evident from Fig.~\ref{fig:f_peak_cold_hot} and Fig.~\ref{fig:f_peak_hot_cold_posterior_total} where $f_{peak}$ posteriors are overlapping for the cold and the warm conditions.

\end{document}